\documentclass[english,twocolumn]{revtex4}

\usepackage[T1]{fontenc}
\usepackage[latin9]{inputenc}
\usepackage{amsthm}
\usepackage{amsmath,mathrsfs,amsfonts}
\usepackage{graphicx}
\usepackage[caption=false]{subfig} 

\makeatletter

\newcommand{\Lie}{\mathcal{L}}
\newcommand{\tl}{\theta_{(\ell)}}
\newcommand{\tn}{\theta_{(n)}}
\newcommand{\tq}{\tilde{q}}
\newcommand{\be}{\begin{equation}}
\newcommand{\ee}{\end{equation}}
\newcommand{\sub}[1]{_{\mbox{\scriptsize{#1}}}}

\makeatother

\usepackage{babel}
\begin{document}

\title{Closest Safe Approach to an Accreting Black Hole}
\author{Benjamin K. Tippett}
\email{ben.tippett@ubc.ca}
\affiliation{Mathematics, Irving K. Barber School of Arts and Sciences, 
University of British Columbia - Okanagan,
3333 University Way,
Kelowna, BC  V1V 1V7, Canada}
\author{Ivan Booth}
\email{ibooth@mun.ca}
\affiliation{Department of Mathematics and Statistics, Memorial University, St. John's, NL A1C 5S7, Canada}

\begin{abstract}
We examine the causal and geometric horizons of dynamical black holes in Lemaitre-Tolman-Bondi collapsing dust spacetimes. Marginally
trapped tubes in these spacetimes may be spacelike, timelike or null and may also be sourced from or disappear into shell-crossing singularities
which we resolve with (timelike) shockwaves. The event horizon kinks when it intersects a shockwave. We calculate the timelike separation 
between the crossable boundary (marginally trapped tubes plus connecting shockwaves) and event horizon. As measured along the crossable 
boundary this function can have discontinuities not only in its derivative but also in the function itself. These features are closely related
to the geometry of the crossable boundary. Finally, we consider the application of this work for future space explorers seeking to make a closest (non-terminal) approach to a black hole horizon. 
\end{abstract}

 %
%
%

\maketitle



\section{Introduction}
\label{Intro}

How close can an observer approach a dynamic black hole without becoming trapped?
From one perspective this is a trivial question: the observer must remain outside the event horizon. However from another it is highly non-trivial:
event horizons are causal in nature and for generic dynamical spacetimes there is no geometric signature that observers can use to determine 
whether or not they have been crossed. The exact location of an event horizon depends on the future history of the spacetime.

%
%
%
%
%

%
%

To humanize this problem, consider the case of a future extreme-astronaut who is seeking to experience the largest tidal-forces ever 
felt a sentient being. Whether she finds herself crushed to death in a singularity or living out her days in leisure, basking in the glory of 
being the hero who made the closest ever approach to a black hole, depends upon whether or not she has crossed the event horizon. How 
can she manage her risks? 

Unfortunately without knowledge of the future, there is no perfect answer. Apparent geometric 
horizons provide a partial answer. If an explorer (or actually a group of explorers since a full surface must be observed) encounters
a marginally or fully trapped surface then this is sufficient to tell her that she is inside the event horizon and has an imminent
date with a singularity\cite{hawking73}. However encountering such a surface is not necessary: the event horizon of a dynamical black hole
can lie well outside any 
marginally trapped surface. As such, probably the best that can be done is to  generate a catalogue of different models of black hole growth, 
and determine their respective closest safe distances; geometrically these are the distances between the event and apparent horizons. 
Then on approaching a black hole, an explorer will at least have a better idea of the risks being taken. 

Less speculatively, knowing the distance betweeen horizons has practical application. In analytical and numerical studies of black hole physics it
is commmon to keep track of horizon geometry and evolution. It is generally assumed (perhaps only tacitly) that the evolution will 
reflect the black hole's interaction with its environment. For example, infalling matter should cause it to grow and incoming gravitational waves 
should distort it. Conversely a distorted black hole horizon should throw off gravitational waves as it oscillates its way back to equilibrium. 

However there is an obvious problem with this picture. Geometric horizons necessarily live inside event horizons and
by construction anything on or inside an event horizon cannot send signals to the outside world. Thus while horizons of any type might 
reflect incoming influences, they cannot be the direct source of signals being transmitted outwards. How then do black holes interact
with their surroundings? The resolution is that a black hole interacts with its surroundings via the gravitational fields \emph{outside} 
the event horizon. This understanding is one of the foundations of the membrane paradigm\cite{Thorne:1986iy}, with the stretched horizon 
standing in for those fields. In many physical situations the stretched, event and apparent horizons will be ``close'' and their evolutions 
 will closely mirror each other. In those cases, apparent or event horizons will serve as proxies for the external fields and tracking 
 the distance between those horizons may serve as an indication of how accurately they reflect the evolution. 

In support of this picture it has been demonstrated that for near-equilbrium black holes, a candidate event horizon may be perturbatively 
constructed near any slowly evolving geometric horizon\cite{Booth:2012xm}. As might be expected, away from equilibrium the situation becomes 
considerably more complicated and the horizons generally separate. Relative horizon locations have been studied in some detail for the simple 
case of Vaidya spacetimes \cite{Booth10,Nielsen11} 
(where the matter accreting onto the black hole is spherically symmetric null dust). In particular, \cite{Booth10} studied their timelike
separation and found that the more densely distributed the matter collapsing into the black hole, the more sharply peaked the timelike
distance profile became.

Lemaitre-Tolman-Bondi (LTB) black hole spacetimes have a much richer physics than that seen in Vaidya. 
For Vaidya spacetimes, null matter shells infall in a very regimented fashion. By contrast, the (still concentric) timelike matter shells of LTB 
are much less disciplined and can spread out, pile up and even overtake each other. This leads to a similarly rich set of possible evolutions 
for the apparent horizon. In Vaidya there is a single apparent horizon which is either null (and in equilibrium) or spacelike (and expanding). 
However, as was shown in \cite{Booth06}, for LTB spacetimes there can be multiple (concentric) marginally trapped surfaces, some of which 
may be timelike (and contracting). Pairs of timelike and spacelike marginally trapped tubes can be created or conversely meet and annihilate. However, in that paper 
event horizons were not considered and parameters were carefully chosen to avoid shell-crossing singularities. 

In this paper we remedy these deficiencies, resolving the singularities with timelike shockwaves and comparing horizon separations for 
spacetimes with the various horizon signatures as well as these shockwaves. We find that while event horizons are
continuous (though not smooth) across shockwaves, marginally trapped tubes are discontinuous:
it is possible for them to appear from and disappear into the shockwave singularities created by crossing matter shells. 
With all these possible behaviours, we need a robust method for finding horizon separations and we
implement and apply such a method. 

To briefly summarize, Section \ref{Hor} reviews horizons: \ref{genHor} is a general discussion of the various horizon types while 
\ref{ssHor} specializes to the mathematics of spherical horizons. Section \ref{sub:Intro-to-tolman} introduces 
general Lemaitre-Tolman-Bondi collapse while sections  \ref{singLTB}  and \ref{sub:Shock-wave-treatment} respectively 
consider singularities and shockwaves in those spacetimes. \ref{LTBH} then discusses horizons and how to locate them. 
Section \ref{compProc} develops our method for computing the distance between horizons and then this method is implemented
and the results considered in \ref{DCSA}. Finally Section \ref{discussion} concludes the paper with a review and discussion of the results.

\section{Geometric and causal horizons}
\label{Hor}

We begin with a more detailed discussion of the various types of horizons followed by a discussion of the geometry and dynamics of spherically 
symmetric marginally trapped tubes. 

\subsection{Types of Horizons}
\label{genHor}

There are a variety of ways to define the boundary of a dynamic black hole \cite{Booth:2005qc,Jaramillo2011}. 
%
%
%
First, there is the event horizon. The classical definition of a black hole is in terms of the causal structure of the full spacetime. 
A \emph{causal black hole} is a part of a spacetime manifold $\mathcal{M}$
which cannot send signals to future null infinity $J^{-}(\mathscr{I}^{+})$. Then, $\mathcal{B}=\mathcal{M}-\mathscr{I}^{-}(\ell^{+})$, is
the interior of the black hole. The \emph{event horizon} is the boundary $\mathcal{H}_{eh}$ between $\mathcal{B}$ and the rest of the spacetime \cite{hawking73}. 
$\mathcal{H}_{eh}$ is traced by a congruence of null geodesics which neither focus to a point nor 
extend out to $\mathscr{I}^+$. 
Though intuitive and mathematically elegant, causal horizons are
inconvenient and not very physical: deciding whether or not a point in $\mathcal{M}$ can send
signals to $\mathscr{I}^+$ will depend on the future development of the spacetime. 

The teleological nature of the event horizon is nicely demonstrated by the most common method used to locate event horizons in numerical relativity
(and the one that we will use in this paper)\cite{Thornburg2007}. 
If a black hole ultimately settles down and stops accreting mass, then it is straightforward to locate the horizon in that final equilibrium regime:
the final state is necessarily Kerr-Newman and so the event horizon is the standard one. One can then find the location of the horizon in the past
by identifying the congruence of null geodesics  that rules the ultimate horizon and ray-tracing them backwards in time. 
Clearly this requires a knowledge of the future evolution of the spacetime (at least up to the point where it reaches equilibrium). 
We return to this in section \ref{LTBH}. 

While the event horizon is an elegant way to demarcate the boundary of the black hole, it is not always so useful when we are attempting to characterize a black hole's evolution. It speaks only to the question of which causal curves are destined to be swallowed, as opposed to identifying strongly 
curved regions of spacetime which can be immediately identified as able to trap light rays.  
Indeed it is possible for event horizons to exist in completely flat 
regions of spacetime where a black hole will, but has not yet, formed (see, for example, the 
discussion in \cite{Booth:2005qc}).

An alternative way to define a black hole is to try to identify the most dramatically curved regions of spacetime. \emph{Trapped surfaces} are a natural
signifier of these regions. These two-surfaces have negative inwards and outwards null expansions  and so the well-known singularity theorems 
guarantee that all of the causal curves that emerge from such surfaces will ultimately
end up in a singularity\cite{penrose65}. Further, if the space time is asymptotically flat, trapped surfaces are guaranteed to be contained in 
$\mathcal{B}$\cite{hawking73}. Thus, given a trapped surface one can unambiguously say that it is part of a black hole without any need to 
explicitly consider the future development of the spacetime \footnote{In fact there are significant complications to this
apparently straightforward idea. Just as it is possible to find event horizons which extend into flat regions of spacetimes, it is also possible to find
trapped surfaces that partially lie in regions of flat spacetime. 
See, for example, the discussions of ``wild'' trapped and marginally 
trapped surfaces and the non-rigidity of marginally trapped tubes 
in papers such as \cite{Schnetter:2005ea,Booth07,BenDov07,Nielsen:2010wq, Bengtsson11}. That 
said, in this paper we restrict our attention to spherically symmetric spacetimes \emph{and} spherically symmetric horizons. With this restriction, 
these complications may be largely ignored. }.  

For spacetimes that have been foliated into spacelike surfaces $\Sigma_t$ (essentially ``instants'' of time) the \emph{trapped} region on a 
given $\Sigma_t$ is the set of all points that lie on any outer trapped surface (in this case only the outward expansion is restricted to be negative). 
Then the boundary between the trapped and non-trapped regions of the hypersurface is a \emph{marginally 
trapped surface (MTS)} (outward expansion vanishes) known as the \emph{apparent horizon}\cite{hawking73}. 
Generalizing from this result, in numerical relativity the term apparent horizon is used slightly differently. There, the apparent horizon is the 
outermost marginally trapped surface found on a given time slice\cite{Thornburg2007}. These are useful both for physical interpretations of 
black hole physics as well as practical numerical reasons: the region inside the apparent horizon is necessarily inside the event horizon and so 
can be excised (along with the singularity) if one is focussed on the external spacetime.  

However obtained, a union of marginally trapped two-surfaces forms a three-dimensional \emph{marginally trapped tube (MTT)} and these are of considerable
interest for both physical and mathematical reasons. Apart from time-evolved apparent horizons, other examples of marginally trapped tubes include 
trapping horizons\cite{Hayward:1993wb} as well as isolated and dynamical horizons\cite{Ashtekar:2004cn}. 

\subsection{Spherical Geometric Horizons}
\label{ssHor}

With this general overview in mind we now focus on spherical MTTs in similarly spherically symmetric spacetimes. These are all we need for this 
study and by restricting our attention in this way we can ignore many complications. The following largely specializes more general results from \cite{Booth07} to spherically symmetric horizons. Similar discussions may be found in \cite{Booth06,Booth10}.

Let $(M,g_{ab})$ be a spherically symmetric spacetime parameterized by coordinates $(t,r,\theta,\phi)$ so that 
\begin{align}
ds^2 = & F dt^2  +  2  G dt dr + H  dr^2 \\
&
 + R^2 \left(d \theta^2 + \sin(\theta)^2 d \phi^2\right) \nonumber\, , 
\end{align}
for some functions $F(\Omega)$, $G(\Omega)$, $H(\Omega)$ where $\Omega = \Omega(r,t)$ is a surface of constant $r$ and $t$. 
$R(\Omega)$ is the areal radius of a given $\Omega$ and it completely defines the intrinsic geometry of those surfaces. 
To wit, denoting the induced spacelike two-metric as $\tilde{q}_{AB}$ they have area element
\begin{equation}
\sqrt{\tilde{q}} = R^2 \sin \theta \, , 
\end{equation} 
and  (two-dimensional) Ricci scalar 
\begin{equation}
\tilde{{R}} = \frac{2}{R^2} \, . 
\end{equation}

 \begin{figure*}
\includegraphics{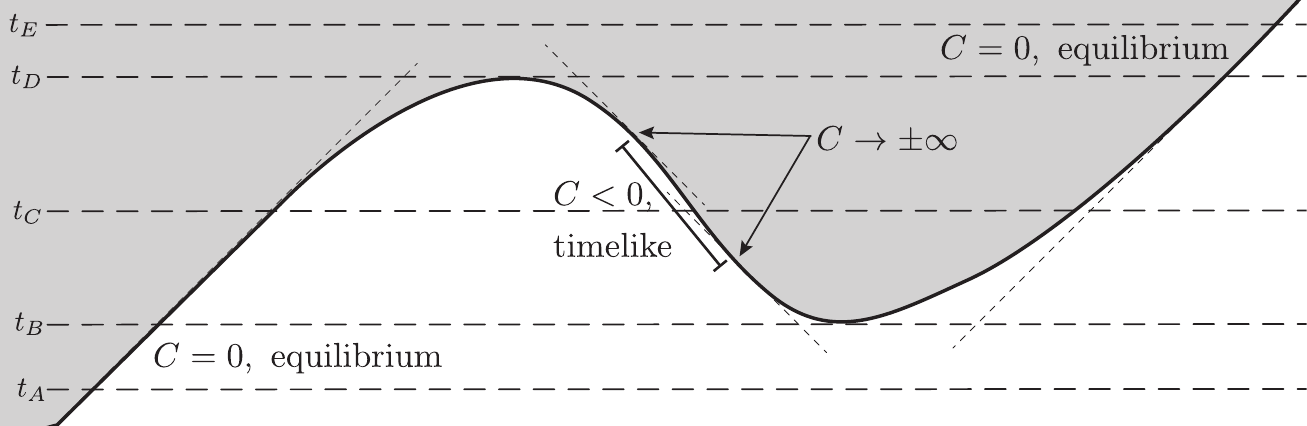}
\caption{Schematic of a smoothly evolving marginally trapped tube that nevertheless exhibits an apparent horizon jump. 
Each point in the diagram represents
a sphere $\Omega(t,r)$. Outgoing and ingoing null rays have slopes $+45^\circ$ and $-45^\circ$ respectively. If the spacetime is foliated into
spacelike ``instants'' labelled by $t$, then at $t_A$ and $t_B$ there is a single MTS. 
At $t_B$ two new MTSs appear, and if one was only tracking the outermost one it would appear to jump. 
At $t_c$  three nested MTSs continue their evolution. Finally two of them meet and annihilate at $t_D$.  
 }
\label{mttCartoon}
\end{figure*}

The normal space to each $\Omega$ has signature $(1+1)$ and so can be spanned by a pair of future-oriented 
null vectors $\ell$ and $n$. Respectively these are outward and inward oriented and we cross-normalize them so that $\ell \cdot n = -1$.  Note that a degree of scaling freedom remains and for any 
function $f$ that shares the spacetime symmetries, $\ell \rightarrow e^f \ell$ and $n \rightarrow e^{-f} n$ continue to satisfy our conditions. 
We assume that this function has been chosen to be spherically symmetric and smooth so that these vectors are defined throughout the spacetime and have those 
same properties. 

The extrinsic geometry of the $\Omega$ may be characterized in terms of the null normals. Given the symmetry, the only non-vanishing components of the 
extrinsic curvature are the  outward and inward expansions:
\begin{equation}
\theta_{(\ell)} =  \frac{2 \Lie_{\ell} R}{R} \; \; \mbox{and} \; \; \theta_{(n)} =  \frac{2 \Lie_{n} R}{R} \, . 
\end{equation}
Since $n$ is inward-pointing we expect $\tn < 0$ (this will be the case for all of our examples) and so classify the $\Omega$ as
\begin{enumerate}
\item \emph{untrapped} if $\tl > 0$
\item \emph{marginally trapped} if $\tl = 0$ and
\item \emph{trapped} if $\tl < 0$. 
\end{enumerate}

Thus, a marginally trapped tube $H$ is foliated by $\Omega$ on which $\tl = 0$. Given a scaling of the null vectors, we can (almost) always find an 
\emph{expansion parameter} $C$ such that 
\begin{equation}
\mathcal{V} = \ell - C n \label{V}
\end{equation}
is tangent to $H$. Then, given $\tl = 0$, the rate of change of the area element up $H$ is 
\begin{equation}
\Lie_{\mathcal{V}} \sqrt{\tq} = \sqrt{\tq} (\tl - C \tn)  = - \sqrt{\tq} C \tn \, . \label{Expansion}
\end{equation}
 Thus with $\tn < 0$ we have the following equivalencies:
\begin{enumerate}
\item $ \mbox{timelike $\mathcal{H}$}  \Leftrightarrow C < 0 \Leftrightarrow \mbox{shrinking MTT}$
\item $\mbox{null $\mathcal{H}$} \Leftrightarrow C = 0 \Leftrightarrow  \mbox{equilibrium MTT}$
\item $\mbox{spacelike $\mathcal{H}$} \Leftrightarrow C > 0 \Leftrightarrow  \mbox{expanding MTT}$
\end{enumerate}
Note the ``almost'': if $H$ becomes parallel to $n$ then $C \rightarrow \pm \infty$. This can occur
but will be dealt with separately as a special case. 

%
%

Finally, it can be shown that for a general spherically symmetric surface 
\begin{align}
\Lie_\mathcal{V} \tl = &  \kappa_{\mathcal{V}} \tl - (\tl^2/2 + G_{ab} \ell^a \ell^b) \nonumber \\
& + C \left(1/R^2 - G_{ab} \ell^a n^b- \tl \tn \right) 
\end{align} 
where $\kappa_{\mathcal{V}} = - \mathcal{V}^a n_b \nabla_a \ell^b$ (which reduces to the surface gravity for Killing horizons) and $R$ 
continues as the areal radius of $\Omega(t,r)$. Thus on a spherical MTT, 
we can solve for $C$ in terms of the surface Ricci scalar and components of the stress-energy tensor
\begin{equation}
C = \frac{G_{ab} \ell^a \ell^b}{1/R^2 - G_{ab} \ell^a n^b} \, . \label{C}
\end{equation}

Key properties of spherically symmetric horizon evolution follow directly from this equation. First, $C=0$ (and $H$ is null and isolated) if and only if
$G_{ab} \ell^a \ell^b = 0$. Physically this is the case where no matter crosses the horizon (there are no gravitational waves in spherically symmetric gravity 
and so the only source of energy for horizon growth is stress-energy). Second, if the null energy condition holds, then $G_{ab} \ell^a \ell^b \geq 0$
and so the sign of $C$ (and signature of $H$) is determined by the relative size of $1/R^2$ and $G_{ab} \ell^a n^b$. 

At least for the case of dust (the focus of our later sections) there is a nice intuitive interpretation for what is happening here
\cite{Booth06}. Dust falling along radial timelike geodesics with tangent $u^a$ with energy density $\rho$ will have stress-energy tensor
\be
T_{ab} = \rho u_a u_b \, . 
\ee
Thanks to the symmetry and radial motion we can always write
\be
u^a = \xi \ell^a + \frac{1}{2 \xi} n^a  \, ,
\ee
for some function $\xi(t,r)$. Then (\ref{C}) becomes
\be
C = \left( \frac{1}{2 \xi^2} \right) \frac{\rho}{1/\mathcal{A} - \rho} \, , \label{CII}
\ee
with $\mathcal{A} = 4 \pi R^2$ the area of $\Omega(t,r)$. Thus, the qualitative geometry of the horizon is determined by the relative size of $1/A$ and $\rho$. 
Now in geometric units,
 density has dimensions of $(1/\mbox{Length})^2$ and in fact in Euclidean space, the density of a sphere with radius $R=2m$ and mass 
$m$ would be 
\be
\rho_{Euc} = \frac{m}{(4\pi/3) R^3} = \frac{3}{2\mathcal{A}}\, . 
\ee
Given these observations, it is natural to view the comparison as being between the effective density of the black hole with that of the matter. 

This interpretation is supported by the observed behaviours of horizons in exact solutions\cite{BenDov:2004gh,Booth06}. 
As shown in Figure \ref{mttCartoon}, $C<0$ can be viewed in two ways.
In the first, there is a single MTT that weaves backwards and forwards through time while always increasing in area. Then $C \rightarrow \pm \infty$
at points where $H$ tips from being spacelike to timelike (or vice versa). From this perspective, increasing the density of the infalling matter causes the area of
the horizon to increase at a greater and greater rate. In all cases the MTT is spacelike and so can be viewed as instantaneous in some reference
frame. However for $\rho > \frac{1}{A}$ the rate of increase becomes so great that it cannot even be viewed as instantaneous in any reference frame: 
instead the MTT has become timelike and travelling backwards in time!

In the second interpretation $C \rightarrow \infty$ is the foliation invariant signature of 
the ``pair creation'' of a spacelike MTT (a dynamical horizon) which expands and a timelike MTT which contracts. 
At those points the infalling matter energy density is sufficient to form a new (outer) horizon, however a region of 
regular untapped spacetime is left inside the new horizon. This is quickly eaten up as: 1) the original (now inner) horizon continues to accrete
matter and expand and 2) the new timelike MTT moves inwards. In the rest frame of the infalling matter this happens with velocity
\be
v = \frac{1 + 2 \xi^2 C}{1-2  \xi^2 C} =  \frac{-1}{2 \mathcal{A} \rho -1}
\ee
and so for $\rho > 1/A$ is always negative: that is it moves inwards more quickly than the matter. 

Note that this behaviour would not be seen in 
a numerical relativity simulation that only tracked outermost horizons: all regions inside the newly formed outer horizon would be discarded and 
so the apparent horizon would be seen to jump outwards. 

Given these complex horizon behaviours, it is natural to consider the situations under which an MTT can begin or end (and so a horizon discontinuously 
appear or disappear)\endnote{See  \cite{Andersson:2005gq,Andersson:2008up} for a more rigorous and detailed discussion of the conditions under which 
an MTT is extendible.}. In spherical symmetry the answer is straightforward: the horizon is inextendible at places where (\ref{C}) is ill-defined. 
Clearly this will happen if either the numerator or denominator become ill-defined. We can exclude the case where the denominator vanishes; 
as noted above at these points the MTT changes signature and apparent discontinuities here arise from our requirement that $\mathcal{V}$ be
of the form (\ref{V}).

Then spherically symmetric MTTs can only be created or destroyed at points where at least part of the stress-energy tensor becomes singular. 
This can happen at the central singularity of a black hole but, as we shall see in Section \ref{Results}, MTTs can also be created or destroyed in 
other singularities. In particular we will see how they can be sourced from shockwave singularities. 

\begin{figure}
\includegraphics{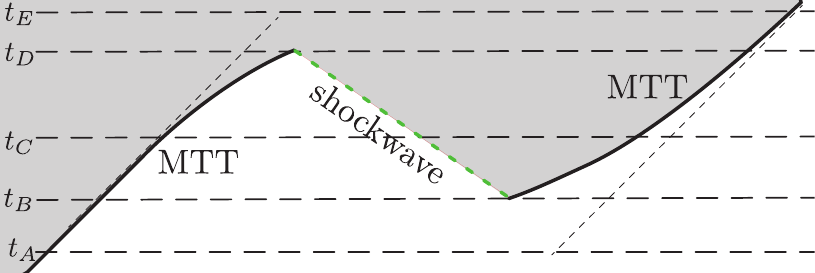}
\caption{Schematic of a discontinuous marginally trapped tube with two pieces connected by a shockwave singularity (dotted in green). Again the 
light gray region is outer trapped.}
\label{Shockwave}
\end{figure}
In the presence of such singularities, the boundary of the spherically symmetric trapped region is not a MTT. Some parts 
of the boundary are marginally trapped, however other sections are instead bound by the shockwave as shown in Figure \ref{Shockwave}. 
Thus we define a new type of boundary world tube. Let us define the \emph{crossable boundary} as being the connected three-surface composed of 
marginally trapped surfaces and shockwaves which is the union of all of the surfaces which an observer can cross to enter a trapped region. It
is this boundary that we will track relative to the event horizon.

\section{Lemaitre-Tolman-Bondi Collapse}
\label{LTB}

We now introduce the  spacetimes which we  will use explore the questions posed in the Introduction. 
Spherically symmetric null dust accreting onto a black hole was already considered in \cite{Booth10}.  Thus we 
move on to the next simplest model: Lemaitre-Tolman-Bondi spacetimes\cite{Bondi47,Lemaitre33,Tolman34},
which can be used to describe the collapse of (still spherically symmetric) timelike dust either to form a new black hole
or grow an existing one. 

\subsection{Introduction to marginal LTB collapse \label{sub:Intro-to-tolman}}

The Lemaitre-Tolman-Bondi (LTB) collapse model provides us with a framework for building analytic models of spherically symmetric gravitational collapse. These models describe spherically symmetric distributions of dust (a pressure-less perfect fluid) free-falling from infinity onto a central point.
We restrict our attention to \emph{marginal} solutions: the matter shells will be marginally bound and so had zero velocity when they started 
their fall from infinity. 

We use (a slightly modified version of) Yodzis' form
of the LTB spacetime metric \cite{Yodzis73}: 
\begin{equation}
ds^{2}=-dt^{2}+ \left(\frac{B(r_o,t) }{A(r_o,t)^{1/3}} \right)^2dr_o^{2}+R(r_o,t)^{2}d\Omega^2\label{eq:metric} \, . 
\end{equation}
The coordinates are: proper time $t$ as measured by the infalling dust, a radial coordinate $r_o$ and the usual spherical coordinates 
$(\theta, \phi)$ so that $d \Omega^2 = d\theta^2 + \sin^2 \theta d \phi^2$ is the metric on a unit two-sphere. The areal radius of 
surfaces of constant $t$ and $r_o$ is:
\be
R(r_o,t)\equiv r_oA(r_o,t)^{2/3}\;, \label{R}
\ee
while the other metric functions are defined in terms of a free function $m=m(r_o)$:
\begin{align}
A(r_o,t)&\equiv1-\frac{3t}{2}\left(\frac{2 m }{r_o^{3}}\right)^{1/2} \; \label{A} \mbox{and}  \\
B(r_o,t)&\equiv1-\frac{3t}{2}\left(\frac{r_o m'}{3 m }\right) \left(\frac{2 m }{r_o^{3}}\right)^{1/2}  \label{B}
\end{align}
where $m' = dm/dr_o$. In particular note that $r_o = R(r_o, 0)$: the areal radius at $t=0$. 

 From these relations (or more fundamentally directly from the Einstein equations) it can be shown that
\begin{align}
R' &= \frac{B}{A^{1/3}} \; \; \mbox{and} \; \; \label{Rprime} \\
 \dot{R} &= - \sqrt{\frac{2m}{R}}  \, , \label{dotR}
\end{align}
where primes and dots respectively indicate partial derivatives with respect to $r_o$ and $t$. Then, in geometrized 
units ($G=c=1$) the associated stress-energy tensor  is:
\be
T_{ab} = \frac{1}{4 \pi R^2} \frac{m'}{R'} [dt]_a \otimes [dt]_b \;  \label{T} \, . 
\ee

This stress energy describes timelike dust that falls along (geodesic) curves of constant $(r_o,\theta,\phi)$ with unit tangent vector
\be
u^a = \left( \frac{\partial}{\partial t} \right)^a \, 
\ee
and we note that the geometric radius of the associated spheres evolves via (\ref{dotR}). 
This is identical with the equation of motion for a test particle falling through a Schwarzschild geometry whose ADM mass is $m(r_o)$. 

Note that observers comoving with the dust see a matter density $\rho = T_{ab} u^a u^b = T_{tt}$. 
Then it is straightforward to see that on a surface $\Sigma_t$ of constant $t$:
\be
m(r_o) = \int_0^{r_o} \mspace{-15mu} \int_0^{\pi} \mspace{-10mu} \int_0^{2 \pi}  \! \! \!  \sqrt{h} \rho \, dr \, d \theta \, d \phi  \, ,
\ee
where $\sqrt{h}$ is the induced volume element. In particular this is true for $t=0$ where $r_o$ is the areal radius and so
$m(r_o)$ can be fruitfully interpreted as the initial mass distribution on $\Sigma_t$. 

%

Thus, the full metric describes the evolution of a spherically symmetric configuration of
dust which free-falls from infinity, focussing upon a central
point. Each sphere of dust follows a path described by constant coordinate
$r_o$ as parameterized by proper time $t$. The areal radius of the shells
as they evolve is $R(r_{0},t)$ and this gives us a direct geometric description of 
how the $r_o$-sphere shrinks as it collapses towards the center. 


\begin{figure}
\includegraphics{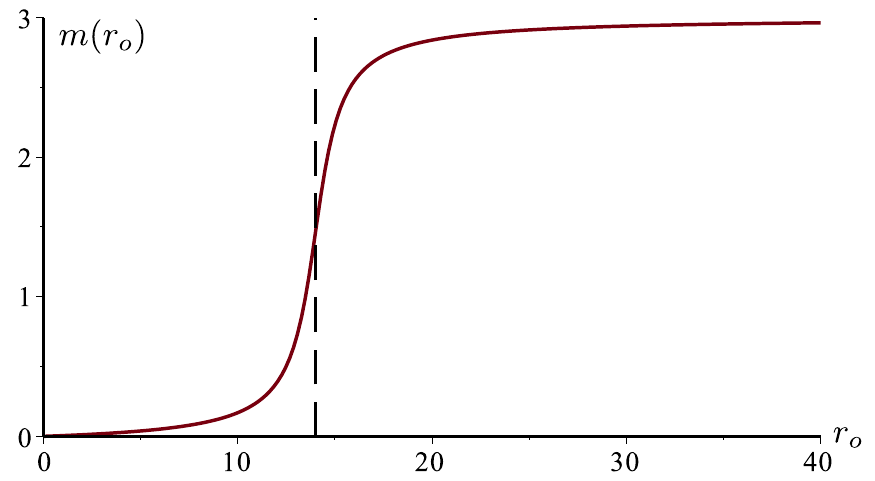}
\caption{The mass function of equation (\ref{massfunction}) with  $m_o = 0$, $\mu =3$, $\bar{r}_o = 14$ and $\Gamma = 1$.}
\label{MassFunction}
\end{figure}

Throughout this section we will demonstrate concepts for the mass function:
\begin{equation}
\begin{split}
m(r_o)  \equiv m_{o} & + \mu \cdot \bigg[
 \frac{\arctan(\Gamma (r_o- \bar{r}_o))}{\pi/2+\arctan(\Gamma \bar{r}_o)} \\ 
& +\frac{\arctan(\Gamma \bar{r}_o)}{\pi/2+\arctan(\Gamma \bar{r}_o )}  \bigg] \\
\end{split} \label{massfunction}
\end{equation}
where $m_o$ and $\mu$ define the limiting values of the mass function ($m(0)=m_o$ and $\lim_{r \rightarrow \infty} = m_o + \mu$), $\bar{r}_o$ 
locates the initial midpoint of the dust shell, and $\Gamma$ determines its initial concentration (increasing $\Gamma$ increases 
the concentration while decreasing $\Gamma$ makes the initial shell more diffuse). 

For definiteness we set 
$m_o = 0$, $\mu =3$, $\bar{r}_o = 14$ and $\Gamma = 1$ (Figure \ref{MassFunction}) and choose to work in units of solar mass. 
Then, physically this choice of parameters corresponds to a spherical shell of three solar masses 
which collapses from infinity and so forms a three solar mass black hole. At $t=0$ the midpoint of the shell has areal radius $14$. 

\subsection{Possible singularities in LTB collapse}
\label{singLTB}

Two different types of singularities can form as the dust collapses: 
these are descriptively named  as \emph{shell focussing} and
\emph{shell crossing singularities} \cite{Yodiz74,Yodzis73} and 
correspond to the two ways in which the stress-energy (\ref{T})
(and Ricci scalar) can diverge.

\subsubsection{Shell Focussing Singularities}

The first possible divergence occurs if 
\be
R(r_o,t\sub{SFS}) \rightarrow 0 \, . 
\ee
Physically this corresponds to the time when an $r_o$-shell encounters the central singularity. By
(\ref{R}) and (\ref{A}) this happens at
\be
t\sub{SFS} = \frac{2}{3}\left(\frac{r_o^{3}}{2m}\right)^{1/2}\;. \label{SFS}
\ee
At $t\sub{SFS}(r_o)$ the areal radius
of the $r_o$ shell goes to zero and so all the motes of that dust shell converge to a central point. 
This is a \emph{shell focussing singularity} (hence the subscript ``SFS'').

%

Note that in the LTB spacetimes that we will be considering, every shell  will eventually encounter the central singularity. 
This follows directly from (\ref{SFS}). The physical range of $t$ associated with each $r_o$ is then bound $t < t\sub{SFS}$. 

\begin{figure} \centering
\includegraphics[scale=0.48]{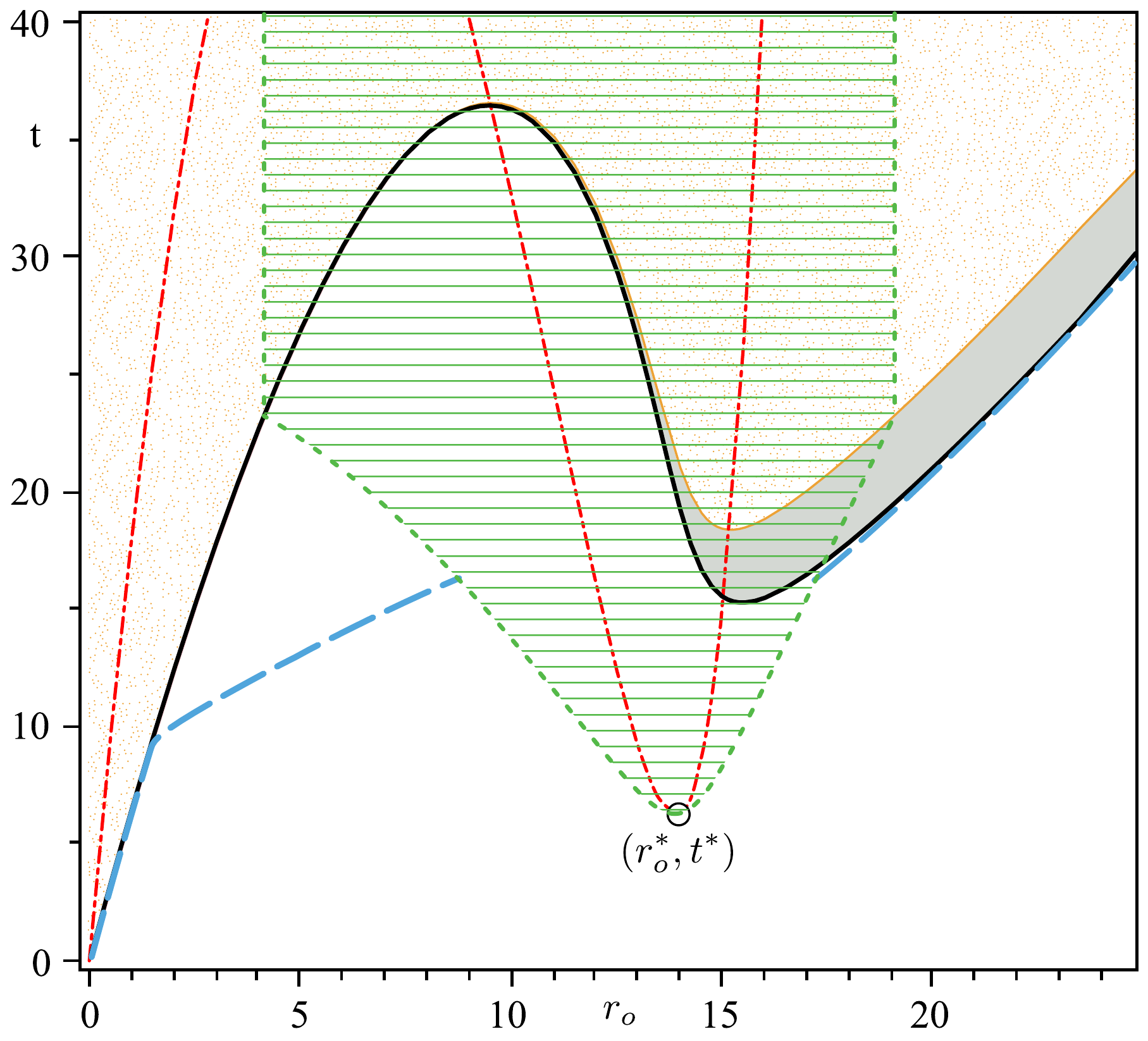}
\caption{The horizons, singularities and shockwave in ($r_o,t$) coordinates for mass function (\ref{massfunction}). The region excluded by
SFS is stippled in yellow with $t=t\sub{SFS}(r_o)$ (\ref{SFS}) the solid yellow line. 
The moment of SCS formation is marked by a white circle while the locus $t=t\sub{SCS}(r_o)$ (\ref{SCS}) is the dash-dotted red line. 
The region between the SFS and marginally trapped tube is shaded grey with the $t=\sub{MTT}(r_o)$ (\ref{MTT}) being the solid black line. 
The dotted green line bounds
the (lined) region that is removed the shockwave excision algorithm: after excision the left and right branches are identified and the resulting 
extrinsic curvature singularity is covered by the thin shell shockwave. Finally, the dashed blue line is the event horizon. 
 \label{fig:shocknaturalcoords}}
\end{figure}

Apart from the physical interest in knowing when a shell hits the singularity, (\ref{SFS}) also gives us a mathematical parameterization of 
the coordinate location of the central singularity. This curve is shown for our sample spacetime with mass function (\ref{massfunction}) 
in Figure \ref{fig:shocknaturalcoords}. Regions of the $(r_o,t)$ coordinate plane beyond the SFS are not part of the spacetime (they 
correspond to regions with negative areal radius: $R<0$). 

\subsubsection{Shell Crossing Singularities}

A second possible divergence occurs if 
\be
R'(r_o,t\sub{SCS})  \rightarrow 0 \; . 
\ee
From (\ref{dotR}) we recognized the trajectories of constant $r_o$ as  mathematically equivalent to the standard geodesics
of a timelike observer falling through a Schwarzschild vacuum with mass $m(r_o)$. However, this depends on the assumption that 
$m(r_o)$ is constant. For some LTB spacetimes (such as those considered in \cite{Booth06}) this is an accurate characterization. 
However in general, geodesics may intersect. In such cases a shell $r_2>r_1$ would cross to the inside of $r_1$, bringing its mass with it. 
Thus the mass contained by $r_1$ would, in principle, increase.  
%
%

In the metric, the signature of such a \emph{shell crossing singularity} (SCS) is $g_{rr} \rightarrow 0$. 
Referring back to (\ref{eq:metric}), an SCS will occur if $B \rightarrow 0$ \emph{and} this happens before 
$A \rightarrow 0$: if the SFS happens first then any potential SCS is irrelevant. Then, from
(\ref{A}) and (\ref{B}), the $r_o$-shell will encounter an SCS if 
\be
m^\prime (r_o) > \frac{3m(r_o)}{r_o} \,  . \label{SCScond}
\ee
Then a shell-crossing singularity forms at: 
\be
t_{\mbox{\scriptsize SCS}}\equiv \frac{\sqrt{2m r_o}}{m^{\prime}} \, . \label{SCS}
\ee

Though intuitively one might expect dust shells to be able to cross without serious difficulties, an examination of the stress-energy tensor  (\ref{T})
shows that, in fact, it diverges at an SCS and so the LTB geometry is not well defined in its causal future. This is particularly serious since 
an SCS can occur at any radius and so is not confined to the interior of the black hole.  As a result, there has been some debate
as to whether their existence constitutes a violation of the cosmic
censorship conjecture \cite{Yodiz74,Yodzis73,Christodoulou84,Newman1986,Kriele94}.
Szekeres showed that the SCSs differ from SFSs, in that the metric at an SCS remains $C^{1}$ (while at an SFS it does not). 
In this sense, SCSs are weak and it was proposed that the cosmic censorship conjecture should be re-worded in order
to accommodate them \cite{Szekeres95}. 

The recognition that the metric at SCS
is still $C^{1}$ suggests an interesting possibility. Thin shells are
well understood, traversable, and are also weak singularities in
that they are similarly only $C^{1}$\cite{Israel66}. It would be convenient if shell
crossing singularities could be interpreted as dynamic thin
shell singularities. However, Goncalves investigated this possibility
and determined that the two are inequivalent \cite{Goncalves02}.

That said, there is a strategy involving thin shells which does allow us to deal with SCSs.
The idea is to surgically excise the region containing SCSs and then identify the 
boundaries of the excised region using the Israel junction conditions: the problematic 
region is removed and replaced with a thin shell\cite{Tegai11}. 

\begin{figure} \centering
\includegraphics[scale=0.45]{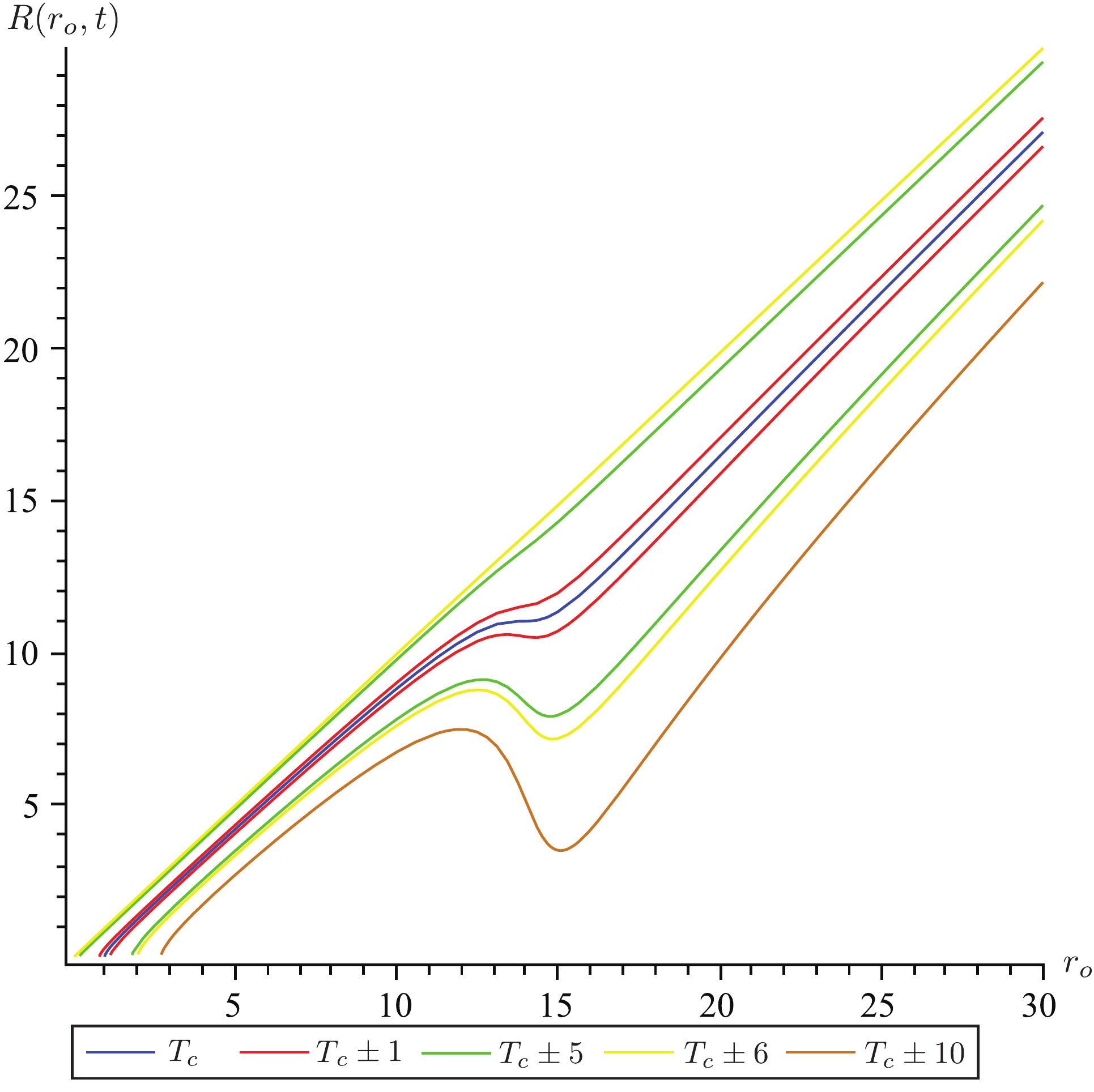}
\caption{$R(r_o,t)$ for various values of $t$. Each curve represents a different value of $t$ 
which increases from the top to the bottom curve. Note that initially $R(r_o,\,0)\equiv r_o$
and $R(r_o,T_c)$ is the middle (blue) curve. 
 \label{fig:Rnomonoto}}
\end{figure}

To motivate the particular form of this procedure that we shall consider, consider how the shell crossing pathologies manifest themselves in 
the $R(r_o,t)$ functions.  As we have seen, a shell of constant coordinate radius $r_o$ is a dynamic object with a similarly dynamic
geometric radius $R(r_o,t)$. Figure \ref{fig:Rnomonoto} plots some of these radii for a particular SCS-forming choice of $m(r_o)$. 
Initially (the topmost curve) $R(r_o,\,0)\equiv r_o$. Curves below that initial one correspond to later times and an SCS forms  
at the critical time $t = T_c$. For $t > T_c$, $R(r_o,t)$ is no longer monotonic in  $r_o$. 

We can then implement an excision in the following way. Start with a 
function $\rho(\tilde{t})$ defined for all $\tilde{t} > T_c$ such that  $R(\tilde{t}, r_o) = \rho(\tilde{t})$
can be solved for multiple $r_o$. For example, on each post-SCS surfaces in Figure \ref{fig:Rnomonoto}, 
there are many ways to find a $\rho$ and $r_{1}<r_{2}<r_{3}\in\mathbb{R}$ such that 
 \[ 
 R(r_{1},\tilde{t})=R(r_{2},\tilde{t})=R(r_{3},\tilde{t})=\rho(\tilde{t}) \, . 
 \] 
For each such $\rho$, one can perform surgery identifying the innermost and outermost roots
and discarding the region in-between. Over this new domain the geometric 
radius is again monotonic in $r_o$.  If we specify
that $\rho$ be continuous, then the surface $R(r_o,\tilde{t})=\rho(\tilde{t})$
becomes the location of a thin-shell, whose existence pre-empts the
formation of SCSs. 

This procedure works to replace the SCS with more palatable thin shells, but clearly is non-unique: there are many ways that we can choose $\rho$. 
We need a well-motivated physical model for choosing a particular $\rho$. In the next subsection we
will review one such proposal which we then use for the rest of this study.

\subsection{Shock Waves \label{sub:Shock-wave-treatment}}

\begin{figure*}
\includegraphics{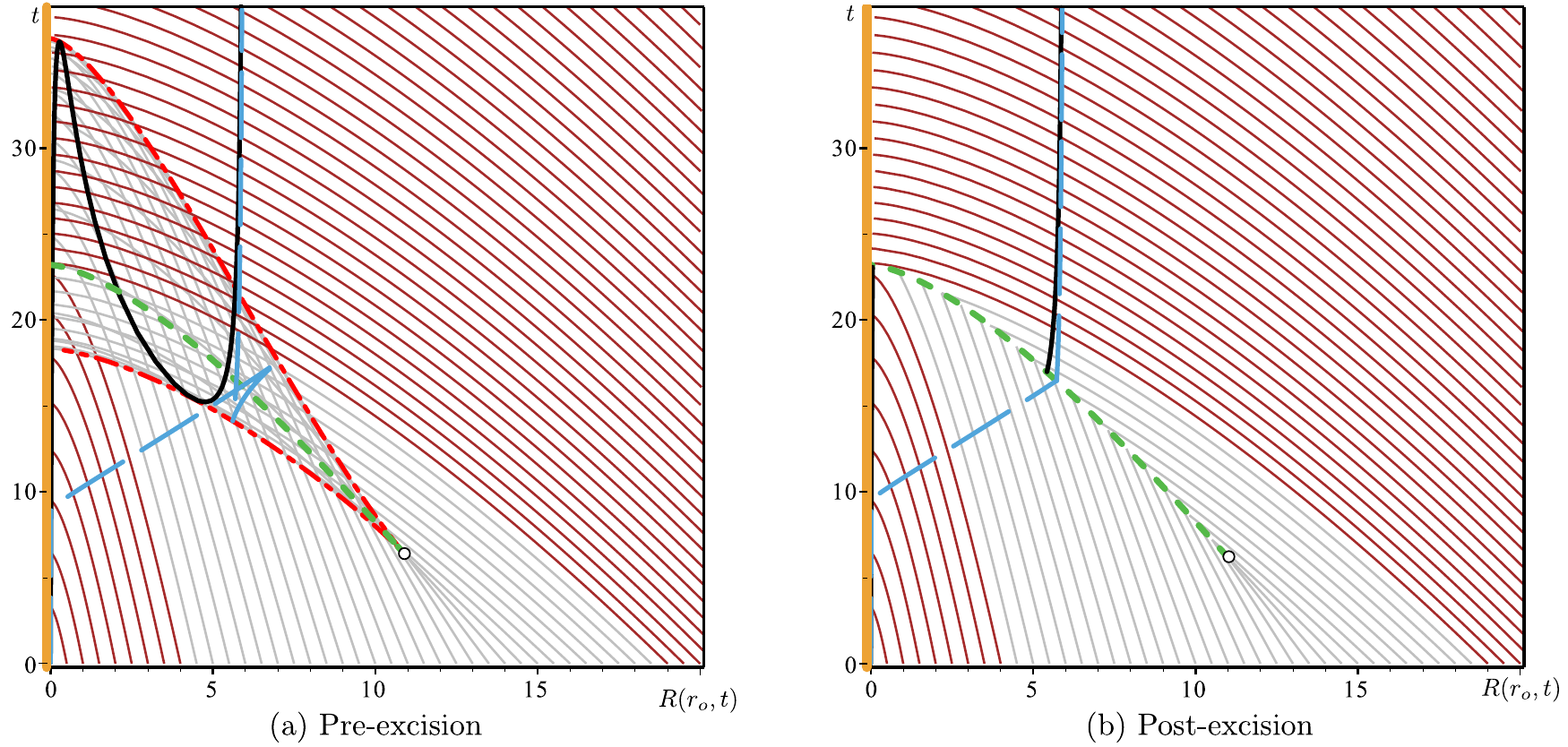}
\caption{Replacing the singular shell crossing region with a shockwave. The colouring scheme of the various quantities is the same as that used in 
 Figure \ref{fig:shocknaturalcoords}. The orange $R=0$ line is the SFS and the red dash-dotted line bounds the SCS region. The MTT is the black
 solid line while the event horizon is the light blue dashed line. Again the onset of SCS is marked with a  white circle and the (future) location of the 
 shockwave is the heavy dashed green line. The remaining lines represent infalling dust shells. The grey ones are terminated on the shockwave
 by the excision process while (post-excision) the dark red ones never encounter the singular region.    }
 \label{shockAsGardening}
\end{figure*}


Nolan suggested that the post-excision thin shell be interpreted as a shockwave generated by the crossing of $r_o$-shells \cite{Nolan03,Lasky07}. 
The physical picture is simple: on the occasion where a slower-moving mote of dust is first overtaken by another more rapidly
moving mote of dust, the dust will stick together. This forms a thin shell of stress-energy whose velocity will be the average of 
the velocities of the two initial motes. As the shock travels inwards, it will accrete more mass, from the inside as it overtakes 
slower moving dust and from the outside as it is overtaken by faster moving dust. 

From this picture one can motivate equations describing a shock (mathematically these follow from the 
Rankine-Hugoniot condition on the system \cite{Nolan03}). In our coordinate system, shock waves are assumed to form at
points $(t^*,r^*_o)$  where 
\be
\left. \frac{\partial R}{\partial r_o} \right|_{(t^*, r^*_o)}= 0 \; . \label{shockInit}
\ee
These are initial conditions. The equations of motion for $R\sub{shock}(t)$, the geometric 
radius of the shock wave at coordinate time $t$, are then 
\be
\frac{d}{dt}R\sub{shock}(t)=\frac{1}{2}\left[ \frac{d}{dt}R(r\sub{left},t)+\frac{d}{dt}R(r\sub{right},t) \right] \, , \label{shockEOM}
\ee
where, as discussed in the previous section $r\sub{left}$ and $r\sub{right}$ are respectively chosen to be the 
smallest and largest values of $r_o$ that satisfy 
\be
R\sub{shock} (t) = R(r_o, t) \; . \label{shockBound}
\ee
At the point of formation $r\sub{left}=r\sub{right}=r^*_o$.

This implementation of this procedure is demonstrated in Figures \ref{fig:shocknaturalcoords} and \ref{shockAsGardening}. 
Figure \ref{fig:shocknaturalcoords} represents the procedure in $(t,r_o)$ coordinates and in those coordinates it tracks the left and right boundaries
of the excision. It is clear that some healthy parts of the spacetime will be excised along with the diseased region: as can be confirmed from Figure 
\ref{fig:Rnomonoto} one cannot simply excise the region bound by $t\sub{SCS}$ as the edges have different areal radii. 

Figure \ref{shockAsGardening} provides an alternative perspective, this time in $(R_o,t)$ coordinates. In \ref{shockAsGardening}(a) the $R(r_o,t)$ 
curves are plotted (keep in mind that $R(r_o,0)=r_o$ and so it is easy to tell which curve is which) and clearly seen to intersect. The effect of implementing the excision procedure is to terminate a subset of curves
(those coloured light grey) at the points corresponding to the excision boundaries 
shown in Figure \ref{fig:shocknaturalcoords}. The location of the future shockwave (and so terminal curve) is shown in (a) and then more appropriately 
in (b). By cropping this subset of $r_o$-curves the shell-crossings disappear.


\subsection{Horizons in LTB spacetimes}
\label{LTBH}

Having developed this understanding of the LTB spacetime, we turn to locating the horizons.
Start with a pair of (cross-normalized) future-oriented radial null vectors
\be
\ell = \frac{\partial}{\partial t} + \frac{A^{1/3}}{B} \frac{\partial}{\partial r_o} \; \; \mbox{and} \; \; n = \frac{1}{2} \left( \frac{\partial}{\partial t} - \frac{A^{1/3}}{B} \frac{\partial}{\partial r_o} \right)
\ee
which are respectively outward- and inward-oriented. The associated expansions are then
\begin{align}
\tl & = \frac{2\left(B \dot{R} + A^{1/3} R' \right)}{BR} \;  \mbox{and} \\
\tn & = \frac{\left(B \dot{R} - A^{1/3} R' \right)}{BR} \, , 
\end{align}
where overdots and primes continue to indicate partial derivatives with respect to $t$ and $r_o$. Applying the definitions of $R$, $A$ and $B$
to eliminate $A$, $B$ and the derivatives of $R$ reduces these to: 
\begin{align}
\tl & = \frac{2}{R} \left(1 - \sqrt{\frac{2m}{R}} \right)\;  \mbox{and} \\
\tn & = - \frac{2}{R} \sqrt{\frac{2m}{R}} \; . 
\end{align}
The outer marginally trapped surfaces are then located at 
\be
\tl = 0  \; \Leftrightarrow \; R = 2m \label{R2m} \; 
\ee
and on those  surfaces the inward expansion
\be
\left. \tn \right|\sub{MTT}  = - \frac{2}{R} =  - \frac{1}{m} \, . 
\ee
These geometric relations will be familiar to anyone who has studied the corresponding problems in Schwarzschild or Vaidya spacetimes. 
Solving (\ref{R2m}) in coordinate form we find that the MTT may be parameterized as
\be
t\sub{MTT} = \frac{2 \left(r_o^{3/2} - (2m)^{3/2} \right)}{3 \sqrt{2m}}  \; . \label{MTT}
\ee
Equivalently, this tells us when the $r_o$-shell crosses the MTT. For our standard example (\ref{massfunction}), these are respectively plotted in
$(t,r_o)$ and $(t,R)$ coordinates as the dashed (blue) curve in Figures  \ref{fig:shocknaturalcoords} and \ref{shockAsGardening}. The 
apparent behaviour of the  MTT inside the region of SCS should not be taken too seriously since the spacetime is not well-defined there. However
post-excision in \ref{shockAsGardening}(b) we see that the MTT does indeed jump along the shockwave singularity as proposed in Figure 
\ref{Shockwave}.

\begin{figure}
\includegraphics{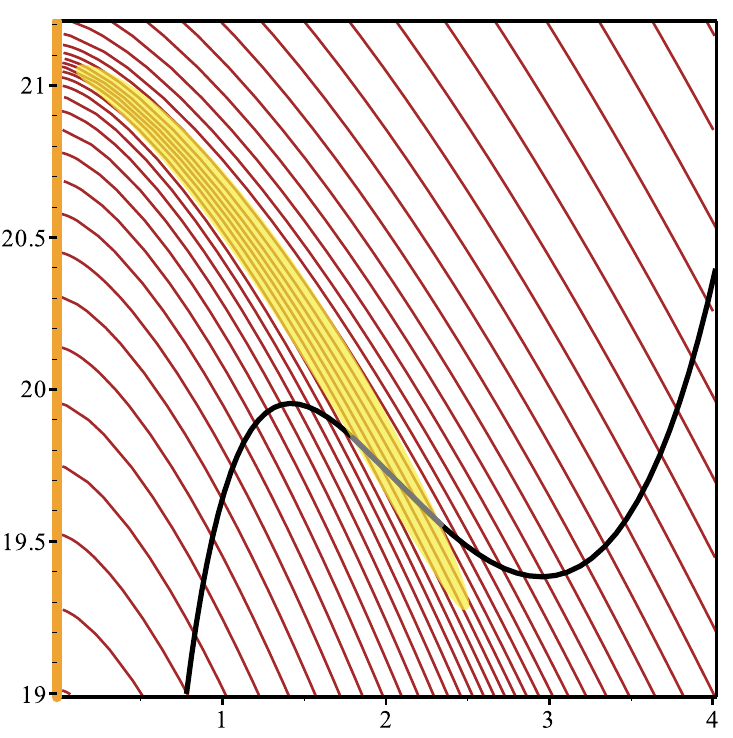}
\caption{An MTT with a timelike section in the $\Gamma =0.25$, SCS-free spacetime.  The critical region where $\rho \geq \frac{1}{\mathcal{A}}$ is coloured in yellow. 
 Where the MTT and critical density region intersect, the MTT becomes timelike. }
\label{FiniteCritical}
\end{figure}

The signature of that MTT can be calculated either directly from (\ref{R2m}) or via (\ref{C}). By either method we find
\be
C = \frac{2m'}{\frac{B}{A^{1/3}}-m'} \; . 
\ee
Then, if the null energy condition holds ($m' \geq 0$): 
\begin{description}
\item{(a)} $\displaystyle m' < B/A^{1/3} \Longleftrightarrow \rho < \frac{1}{\mathcal{A}} \Longleftrightarrow $  MTT is spacelike
\item{(b)} $\displaystyle m' > B/A^{1/3} \Longleftrightarrow \rho > \frac{1}{\mathcal{A}} \Longleftrightarrow $  MTT is timelike \, . 
\end{description}
A lot is happening in Figures \ref{fig:shocknaturalcoords} and \ref{shockAsGardening}, so to clearly see spacelike--timelike transitions we turn to a simpler, 
shockwave-free example. Figure \ref{FiniteCritical} uses the same parameters as our standard example except that we set $\Gamma = 0.25$. In this case 
shells do not cross before the SFS and so the spacetime is shockwave free. The figure shows the region where the density exceeds
the critical density ($\rho \geq \frac{1}{\mathcal{A}}$). Where the MTT intersects that region it is timelike.  Note that matter shells may both enter and 
leave the critical region: regions of high density may both form and disperse. Further, in the timelike

This takes care of the MTT. It remains to locate the event horizon. This will be an outward oriented null geodesic and so a solution of 
\be
\frac{dt\sub{EH}}{dr_o} = \frac{B}{A^{1/3}} \, . \label{EHeq}
\ee
As a boundary condition we assume that the horizon ultimately asymptotes
into an equilibrium state $R_{EH}  \rightarrow R_o$ for some constant $R_o$. Then it is simply a matter of integrating backwards in 
time from that final state. Note any minor errors in the boundary condition (for example integrating from a time when it hasn't quite settled
down) will not be significant as this method exponentially converges to the true event horizon \cite{Thornburg2007, Booth10}. 

The event horizon is plotted by these methods in Figures \ref{fig:shocknaturalcoords} and \ref{shockAsGardening}. Again the apparent contortions
inside the SCS in \ref{shockAsGardening}(a) are symptomatic of spacetime not being well-defined in that region. However in 
\ref{shockAsGardening}(b) the event horizon is well-defined and continuous. This helps to highlight the qualitatively different definitions of these
two objects: the MTT is defined by local geometric quantities while the event horizon is retroactively constructed as the null surface that will
ultimately asymptote to $R_o$. It is continuous by construction: one evolves back from the final state to the shockwave and then picks up the 
evolution again at the same location though on the other side. However there is a kink: ${d R\sub{EH}}/{dt}$ is not continuous. 

\section{Closest safe distance during marginal LTB Collapse}
\label{Results}

We now understand how to replace shell crossing singularities with more tractable shockwaves and have seen how those shockwaves 
can affect event horizons and MTTs. In these spacetimes the crossable boundary is no longer an MTT but instead is made up of 
both MTT and shockwave segments. 
Next we would like to characterize the timelike separation between that crossable boundary and the event horizon. For each point on the crossable
boundary we will find the maximum timelike geodesic length from that point to the event horizon. 

For definiteness we restrict our attention to LTB spacetimes in which a finite mass spherical (thick) shell of dust falls onto a 
pre-existing black hole as parameterized by the mass function (\ref{massfunction}) -- though with a different set of mass parameters from those
considered in earlier examples.



\subsection{Computational procedure}
\label{compProc}

\begin{figure*} \centering
 \subfloat[Spacelike MTT: Normal geodesic intersects event horizon]{  \label{fig:Xspacelike}  \includegraphics[trim=0mm 0mm 0mm 0mm,clip,width=0.4\textwidth]{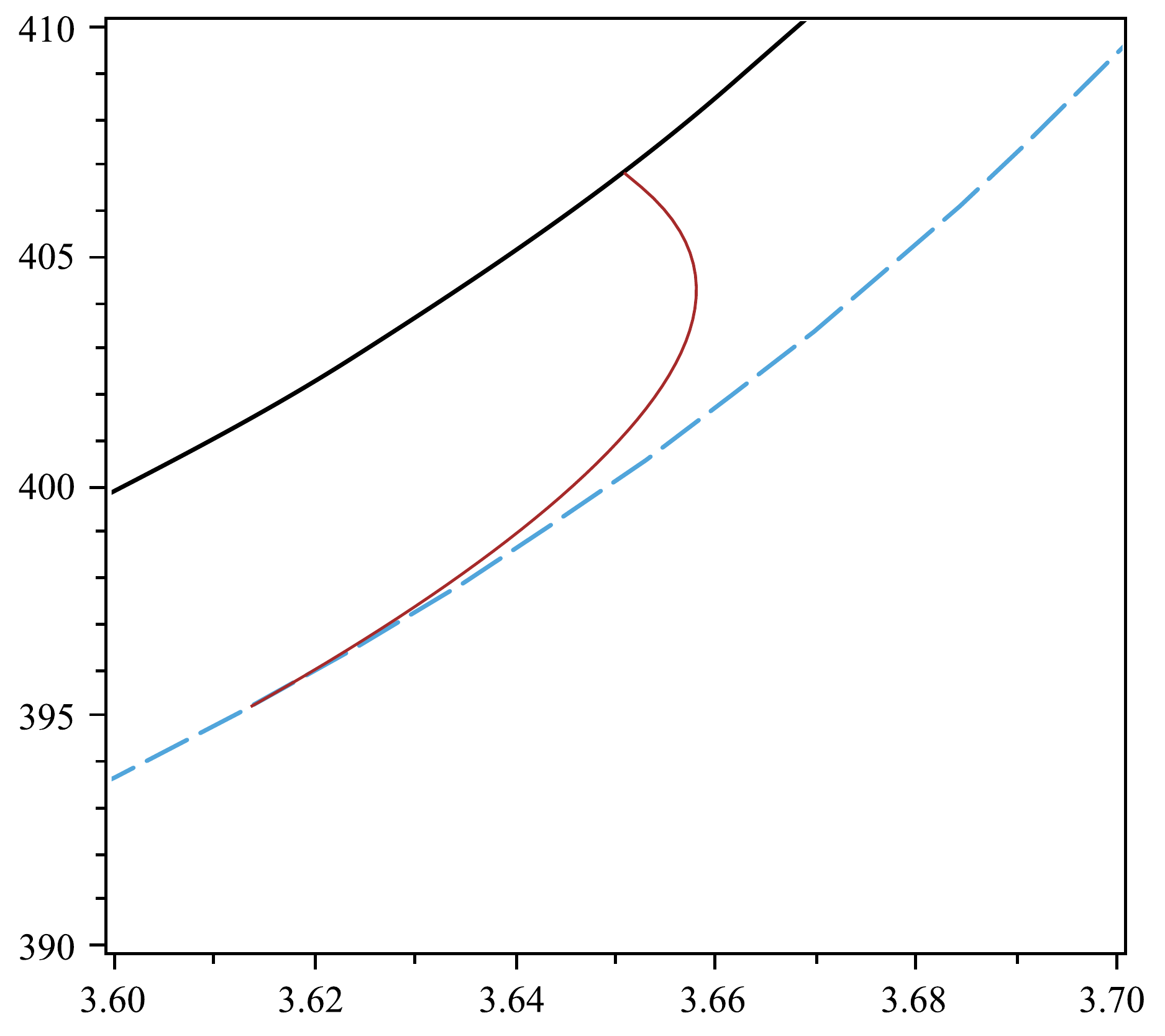} }
 \subfloat[Spacelike MTT: Normal geodesic doesn't intersect event horizon]{  \label{fig:GeoDontX}  \includegraphics[trim=0mm 0mm 0mm 0mm,clip,width=0.4\textwidth]{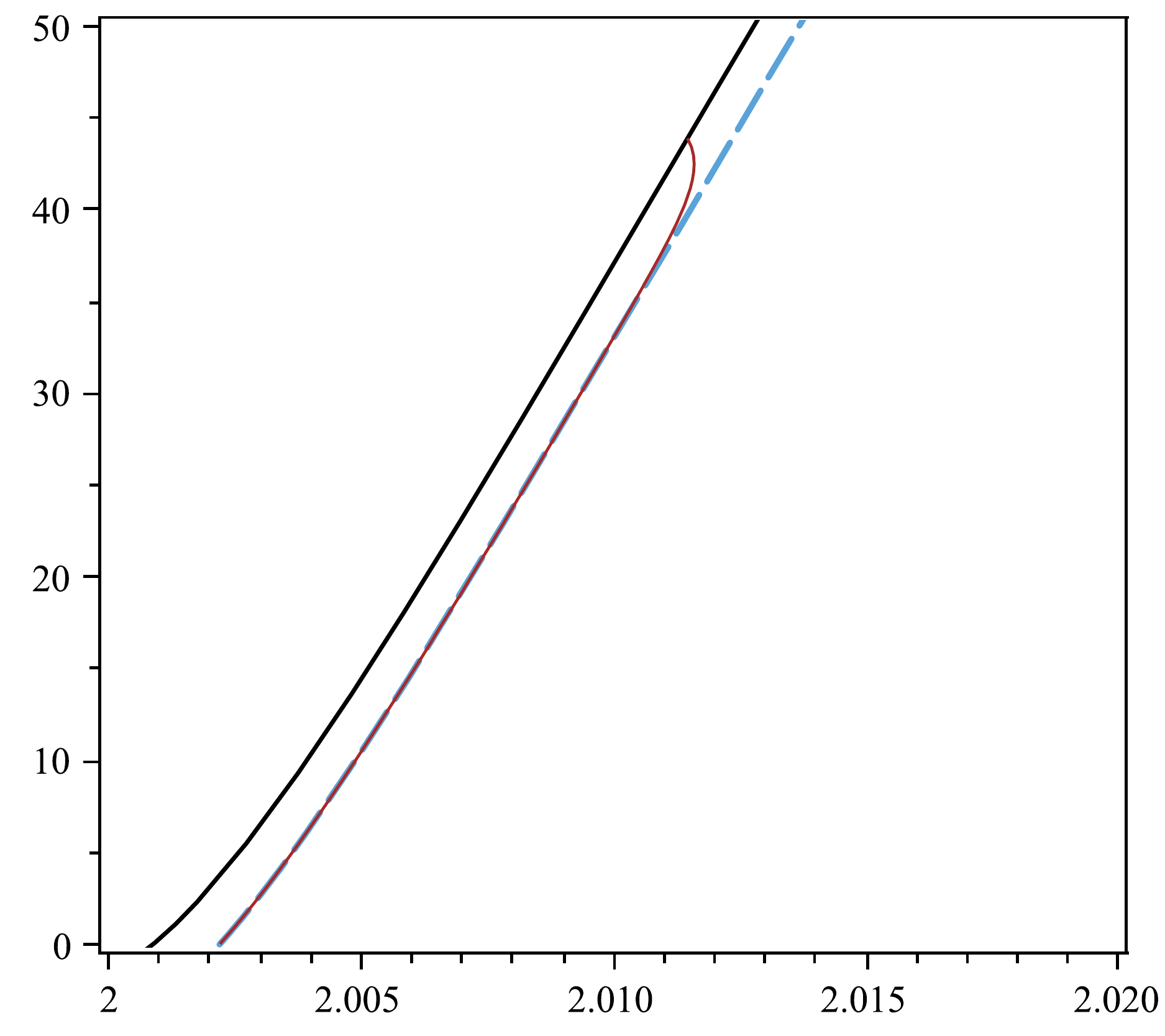} }
 
  \subfloat[Crossable boundary (in this case shockwave) is not spacelike. Several timelike geodesics shown.]{  \label{fig:splinemaker}  \includegraphics[trim=0mm 0mm 0mm 0mm,clip,width=0.4\textwidth]{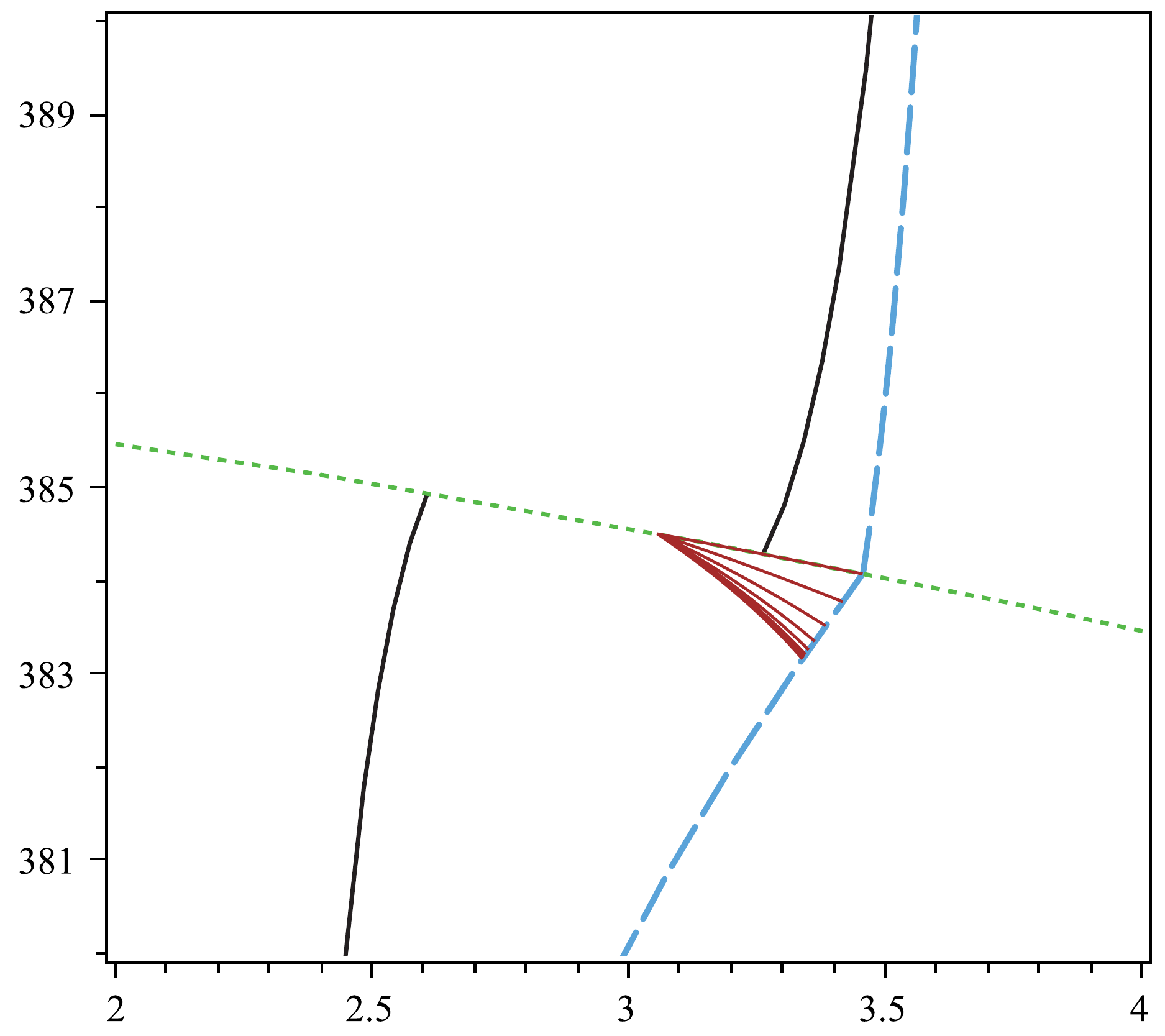} } 
  \subfloat[Spacelike MTT: normal geodesic crosses shockwave]{  \label{ShockR}  \includegraphics[trim=0mm 0mm 0mm 0mm,clip,width=0.4\textwidth]{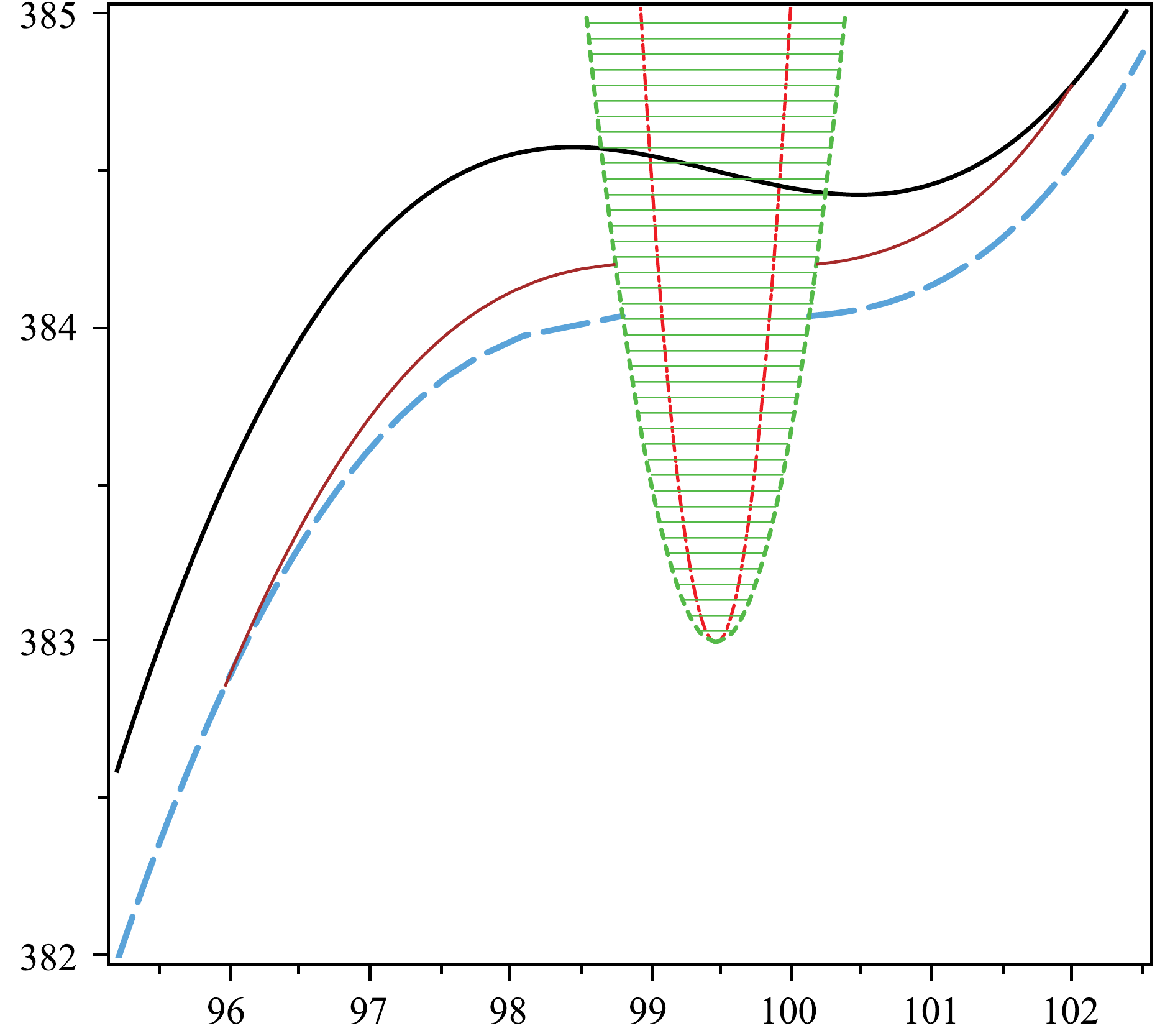} }

      \caption{The different possibilities arising when defining distance between horizons. As in previous diagrams, MTTs are black solid lines, event horizons are dashed light blue lines, shockwaves are dotted green and SCS regions are shaded green. The timelike geodesics used in measuring distance between horizons are solid dark red lines.  }
      \label{Geodesics}
\end{figure*}

Qualitatively, the possible crossable boundaries and corresponding initial conditions  classified by their most dramatic behaviours are:
\begin{enumerate}
\item \textit{Regular expansion:} If the initial distribution of mass is fairly spread-out, the crossable boundary will be an MTT which grows
smoothly and is spacelike throughout its period of expansion. For Vaidya spacetimes this is the only possible behaviour and was studied in detail in \cite{Booth10}. 
\item \textit{Expansion with ``jumps'':} As the initial distribution of mass becomes more concentrated, it is possible for new horizons to 
form around pre-existing ones: the apparent horizon ``jumps''. Equivalently, the MTT includes both spacelike and timelike sections\footnote{Note
that it is also possible to have apparent horizon jumps for during purely spacelike evolutions and in fact many horizon jumps observed in 
numerical relativity are of just this type (see, for example \cite{Ziprick:2008cy,Chu:2010yu}). In these cases
the jumps may be eliminated by an appropriate refoliation of the spacetime. However, if the MTT tips into begin timelike then the jump will be 
present in all possible foliations of the spacetime. }. 
This type of evolution was studied for LTB spacetimes in \cite{Booth06} however the separation function between the MTT and event horizon was not considered. 
\item \textit{Expansion with SCS:} As we have seen, for a sufficiently concentrated initial distribution, the mass shells will cross
and singularities result.  Following the procedure outlined in Section \ref{sub:Shock-wave-treatment}, we can remove the SCS and replace it with a 
shockwave, which may form part of the crossable boundary. 
\end{enumerate}

If the crossable boundary is an always spacelike MTT, then there are theorems that help us to define the separation from the event horizon. 
Recall (\cite{wald}, Theorem 9.3.5) that for a spacelike hypersurface $\Sigma$ and a point $P$ 
not on that surface  the maximum timelike distance between $P$ and
$\Sigma$ is found along the timelike geodesic orthogonal to $\Sigma$ that also passes through $P$ (assuming no conjugate points).
Thus for a spacelike MTT it is straightforward to find maximum times between horizons. One simply 
constructs timelike geodesics normal to the MTT, finds where they intersect the event horizon and measures their length. For the cases studied
in \cite{Booth10} there were no conjugate points between the horizons and so the normal geodesics did give extremal distances. 
A typical MTT-orthogonal timelike horizon-to-horizon geodesic is shown in Figure \ref{Geodesics}(a). 

However, there is an extra wrinkle. In that paper it was also observed that, in general, there are points on the 
event horizon that do not lie on any MTT-orthogonal geodesic. This was shown to be generic for 
a Vaidya transition between equilibrium states. From the time when the infalling matter first hits the MTT until some later critical time partway
through the evolution, the geodesics asymptote to rather than intersect the event horizon (as they travel into the past). However, in that approach 
they are ``nearly null'' and so remain of finite length. That length approaches zero as the MTT approaches equilibrium (again in the past).  
An example of such a geodesic is shown in Figure \ref{Geodesics}(b).

In these cases each point on the event horizon is intersected
by an MTT-orthogonal geodesic whose length gives the maximum timelike distance from that point to the MTT. 
However there are also the remaining MTT-orthogonal geodesics which do not intersect the event horizon: the points where those originate are not endpoints of any 
event-horizon-originating curve of maximum proper length. However, they still provide an upper bound on timelike distance from the event horizon to the MTT from 
those points and so it is useful to calculate their length. Numerically we do this by truncating their evolution and recording the 
length of the truncated geodesic. With an appropriate choice of truncation time, 
the numerical error can be shown to be insignificant. For an extended discussion of this point see \cite{Booth10}.  


%
%
%

For the cases where the crossable boundary contains timelike sections (either MTT or shockwave) the maximum distance theorem no longer applies.
We certainly wish to include these cases in our study and so resort to a more robust, though less mathematically elegant, strategy for quantifying the
distance. Given a point on the crossable boundary, we generate a family of geodesics originating at that point as shown in Figure \ref{Geodesics}(c).
We find the intersection of each of these with the event horizon and then the length to that point (or apply the truncation procedure considered above
if there is no intersection). These lengths are then used to generate a spline, which in turn is used to generate further test geodesics in the 
neighbourhood of the maximum length which is then used to generate a more refined spline. In this manner the geodesic spanning the
maximum distance can be determined and it is this length that will serve as our definition for the \emph{distance between} the crossable boundary 
and the event horizon. 

%
%
%
%

For spacelike MTTs this longest geodesic will be the MTT-orthogonal one unless the geodesic contains either a conjugate point or intersects the 
shockwave. Conjugate points may change which geodesic is longest but are not a problem for our new method: as long as the spacetime is regular
it will work. However, shockwaves are (mild) singular surfaces and so we need rules for how they refract timelike geodesics. 
Recall that while the location of the shock at any time has a unique value in
the geometric $R(r_o,t)$ coordinates, it is represented as two timelike three-surfaces in the pre-excision $(r_o,\, t)$ coordinates, Figure \ref{Geodesics}(d). Intuitively 
the refraction corresponds to the fact that, at any point in time, there are always two families of inequivalent geodesics hitting the shock: those
from the interior and those from the exterior. In the spirit of the Israel-Darmois junction conditions and following standard practice, we identify these 
by demanding that geodesics project into the shockwave hypersurface in the same way when they enter and exit. 
This requirement fully determines how geodesics refract as they cross the shockwave.



%
%
%

\subsection{Distance of closest safe approach}
\label{DCSA}

\begin{figure*}
\includegraphics{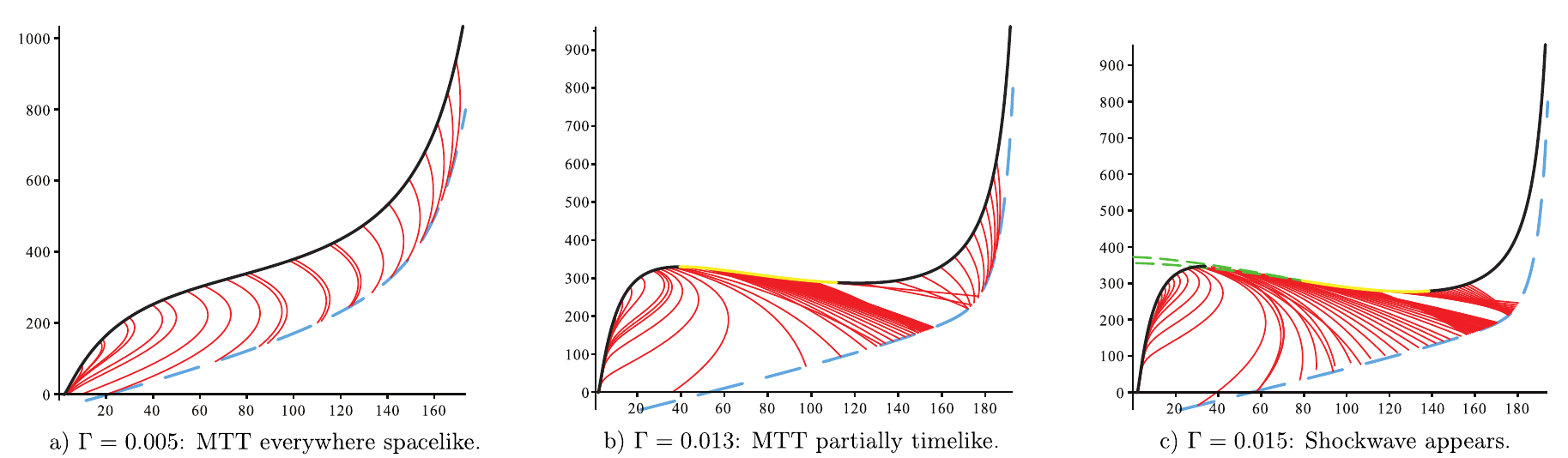}
\caption{The event horizon (dashed blue) and crossable boundary (black when spacelike, yellow when timelike) for a range of values of the concentration parameter 
$\Gamma$. As usual regions of shell-crossing singularities are outlined in dashed green lines. Longest geodesics from the crossable boundary are plotted in red. }
\label{Range}
\end{figure*}

We now apply these procedures  to study horizon separations for a sample set of spacetimes. 
As before the mass function is given by (\ref{massfunction}):
\begin{equation}
\begin{split}
m(r_o)  \equiv m_{o} & + \mu \cdot \bigg[
 \frac{\arctan(\Gamma (r_o- \bar{r}_o))}{\pi/2+\arctan(\Gamma \bar{r}_o)} \\ 
& +\frac{\arctan(\Gamma \bar{r}_o)}{\pi/2+\arctan(\Gamma \bar{r}_o )}  \bigg] \\ \nonumber
\end{split} \end{equation}
however this time we set $m_o =1$, $\mu = 100$ and $\bar{r}_o = 300$. 
Again we work in solar masses, so physically these solutions describe a
very large $100M_\odot$ shell accreting onto an $1M_\odot$ black hole. 
At $t=0$, the middle of the shell has areal radius $300M_\odot$and  will be moving inwards with velocity given by (\ref{dotR}).  
The concentration of that shell around $r_o = 300M_\odot$ is determined by $\Gamma$.

Tuning $\Gamma$ (and so the initial density of the matter) qualitatively changes the crossable boundary as shown in Figure \ref{Range}. 
For $\Gamma \lesssim 0.012$ the boundary is an MTT that is everywhere spacelike. At two key values the behaviour changes. 
At $\Gamma \approx 0.013$ the matter becomes concentrated enough to force the MTT to develop a timelike region: essentially this signals 
a new black hole forming around the old one. Then at $\Gamma \approx 0.015$ the matter shells begin to overtake each other outside the 
MTT and a shell-crossing singularity appears which, as usual, we excise and replace with a  shockwave.  

\begin{figure*}
\includegraphics{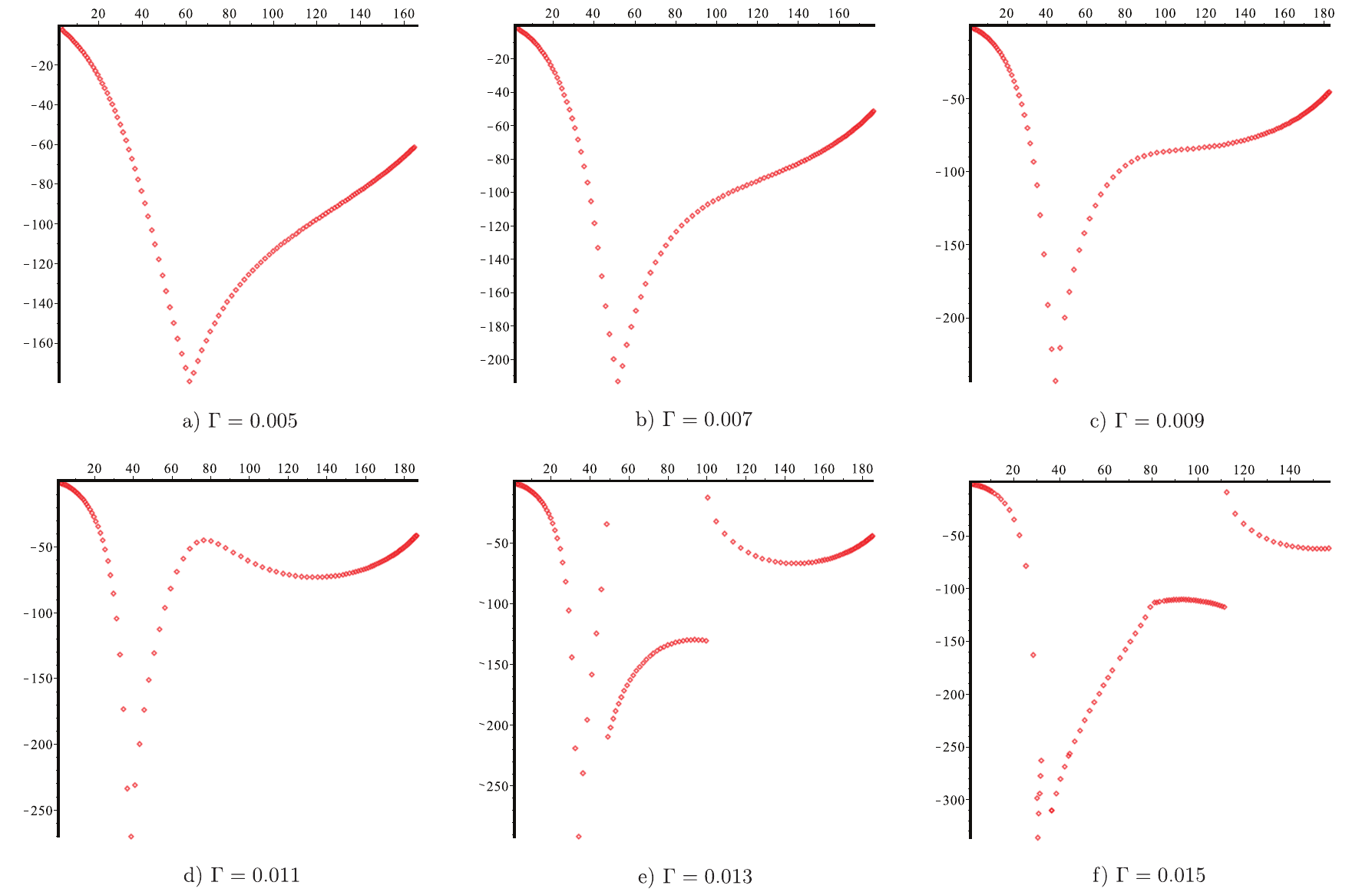}
\caption{Maximum timelike separation from the  crossable boundary to the event horizon for several values of the concentration parameter $\Gamma$. Because the crossable boundary always lies in the future of the event horizon, all values are negative. Note that these graphs are upside-down as compared to the ones in  \cite{Booth10}: in that paper the graphs showed distance from the event horizon to the crossable boundary whereas these show the opposite. }
\label{Distances}
\end{figure*}


Using the techniques outlined in \ref{compProc} we calculated the distances between the event horizon and crossable boundary for the  
cases displayed in Figure \ref{Range} along with some intermediate values of $\Gamma$. These distances are shown in Figure \ref{Distances}
and qualitative  features of those graphs reflect the behaviours of the horizons. 

To begin, note the initial (downward) spike at $R_\star \approx 40-60$ in all of those graphs. This feature is familiar from Vaidya
 and (as discussed in previous section) signals the transition from when
the longest timelike geodesics from the MTT do not intersect the event horizon ($R < R_\star$) to when the do ($R>R_\star$). 
As in the Vaidya case,  this switch moves to earlier times as the initial matter concentration and maximum separation increase. 
Further, as in that case the maximum separation is at most two to three times the size of the final mass. 

However, even for these purely spacelike horizons, the separation distance graphs in Figure \ref{Distances}a)--d) have a more complicated structure
than they did for Vaidya. In contrast to Vaidya (Figure 6 of \cite{Booth10}), the separation graphs are no longer (approximately) symmetric:  for LTB they 
develop a ``bump'' which peaks at $R \approx 130$. Ultimately at  $\Gamma \approx 0.013$ the left slope of this bump evolves into a discontinuity at 
$R \approx 100$ in Figure \ref{Distances}e). This signals the development of a timelike section in the MTT and the consequent switch from 
spacelike-hypersurface orthogonal geodesics to a manual search for longest geodesics. 
This change of behaviour can also be seen in Figure \ref{Range}b) as geodesics originating from $R \approx 100$ begin to cross those
originating at later times along the MTT: points on the event horizon may be the endpoints of multiple locally-longest from the MTT. 

The appearance of the shockwave outside of the MTT at $\Gamma \approx 0.015$ generates a new qualitative change in the distance graphs 
in \ref{Distances}f). In that figure from $R \approx 40-80$ an approximately straight line replaces the more smoothly curved distance variation 
characteristic of smaller values of $\Gamma$ and represents the geodesics originating from the shockwave section of the crossable boundary. 
Note however, that this figure also still contains a timelike MTT section as signalled by the discontinuity for $R \approx 110$ and corresponding
geodesic crossing in \ref{Range}c). 

Apart from this close correspondence between the qualitative behaviour of the horizons and features in the distance graphs we can also consider the 
values of the maximum distances between the causal and geometric horizons.  As for Vaidya, that maximum separation was from the critical point where 
maximum length geodesics transition from not-intersecting to intersecting the event horizon. Further, again in correspondence with Vaidya, that maximum 
separation was on the order of a few times the final mass -- in our examples around $340$ as compared to  $m_o + \mu = 101$.

\section{Discussion}
\label{discussion}

We have examined horizon evolutions in general LTB spacetimes. For sparse initial dust distributions the MTT remains null (in isolation) or spacelike
throughout its evolution while increasing the initial concentration can cause a transition to timelike evolution. Physically this corresponds to the formation of a new
black hole outside of the original one. After this formation, the orphaned bubble of regular spacetime is  rapidly consumed by a spacelike branch of the MTT 
expanding from the original horizon and a timelike branch contracting from the new horizon. This timelike branch contracts faster than the infalling matter shells. 

These behaviours have been seen in earlier studies however in this work we have allowed for even more concentrated initial matter distributions which 
generate shell crossing singularities. Following existing procedures these were excised and reinterpreted as shockwaves. We then studied boundary 
evolutions in these most-general LTB spacetimes and saw how the MTT may develop discontinuities as it can be generated from or terminate on a shockwave. 
By construction the event horizon remains continuous across a shockwave however it kinks when it crosses a shockwave: its first derivative is no longer continuous. 

With these more general horizon behaviours, the maximum timelike separation between the crossable boundary and event horizon becomes similarly more
complicated. In contrast to the Vaidya case, it is no longer possible to rely on distance theorems to automatically generate longest possible geodesics from the 
MTT: these are not valid when the MTT becomes timelike and with our more complicated geometries conjugate are no longer ruled out. Instead we explicitly 
searched for longest timelike geodesics and constructed them for a range of parameter values. The separation graphs had qualitative features that could be 
matched against the MTT behaviours. In addition to the standard non-EH-intersect/EH-intersect transition they develop a discontinuity to mark the appearance
of timelike sections of the MTT and a first-derivative discontinuity when the geodesics must cross the shockwave. 

In the examples that we considered, the maximum separation between the MTT and EH was 3--4 times the maximum horizon mass and that maximum was 
located at the non-EH-intersect/EH-intersect transition. However we have not run enough tests to be say with confidence whether or not this is a general result. 
In the presence of timelike sections and shockwaves, the search for longest geodesics was computationally intensive and so we have only considered a small
section of the total possible parameter space of solutions. For example it should be possible to construct LTB spacetimes with arbitrary numbers of shockwaves 
and it might even be possible to have them intersect each other. For such extreme spacetimes it is quite likely that the timelike separation graphs would become
similarly complicated and could easily develop extreme behaviours. We leave any investigation of such spacetimes to future investigations. 

\begin{acknowledgements}
 This work was supported by the Natural Sciences and Engineering Research Council of Canada. 
\end{acknowledgements}

\bibliographystyle{unsrt}

\bibliography{paper}

\end{document}